\documentclass[apl,twocolumn,amsmath,amssymb]{revtex4-1}
\usepackage{bm}
\usepackage{graphicx}
\usepackage{color}
\draft 

\begin{document}


\title{
\textcolor{blue}{Intense terahertz emission from intrinsic Josephson junctions by external heat control}
}




\author{Hidehiro Asai}
\author{Shiro Kawabata}

\affiliation{Electronics and Photonics Research Institute (ESPRIT), National Institute of Advanced Industrial Science and Technology (AIST), Tsukuba, Ibaraki 305-8568, Japan}


\date{\today}

\begin{abstract}
A practical method for realizing intense terahertz (THz) emission from intrinsic Josephson junctions (IJJs) by utilizing external local-heating is proposed and demonstrated theoretically.
An artificial temperature distribution induced by local heating strongly excites Josephson plasma waves inside IJJs.
Accordingly, the emission power of the THz wave is enhanced drastically, and it can reach the order of mW.
Our result indicates that the use of local heat control is a powerful method to realize practical solid-state THz-emitters based on IJJs.
\end{abstract}

\maketitle 

The use and manipulation of terahertz (THz) radiation has attracted considerable interest because of its tremendous potential in  technological applications such as the non-destructive inspection of materials, medical diagnosis, bio-sensing, and high-speed wireless-communication.
THz-wave generation is fundamental to these applications, and  various emitters such as quantum-cascade lasers and resonant-tunneling diodes have been studied thus far.\cite{tonouchi,qcl}

Subsequent to the experimental realization of strong THz emission from a $\textrm{Bi}_2\textrm{Sr}_2\textrm{CaCu}_2\textrm{O}_{8+\delta}$ (Bi2212) single crystal~\cite{Ozyuzer}, high-$T_{\rm c}$ superconductors have also been considered as promising candidates for compact solid-state THz-sources.  
THz waves from a Bi2212 is induced by an AC Josephson current flowing through its layered structure that is referred to as an intrinsic Josephson junctions (IJJs); IJJs comprise natural stacks of Josephson junctions composed of superconducting $\textrm{Cu}\textrm{O}_{2}$ and insulating layers.
%
In the last few years, a large number of studies have been carried out on THz emission from mesa-structured IJJs both experimentally\cite{Ozyuzer,jpsjKado,hotWang,hot2Wang,aplKakeya,An,TBenseman,Sekimoto,TMinami} and theoretically.\cite{inphaseKoshelev,kinkHu1,fdtdKoyama1,FukuyaTachiki,Klemm,reviewSavelev,inphaseAsai,hotGross2}
Although these IJJ emitters are able to cover the frequency range of $0.3$--$1$ THz, the observed emission powers of the order of $30$ $\mu$W are considerably lower than 1 mW that is required for practical applications.
Therefore, further investigations toward the realization of high-power emission are important.

In a recent study, a $hot$ $spot$ in a mesa wherein the temperature is locally higher than the superconducting critical temperature $T_c$ during the THz emission process has been observed.\cite{hot2Wang,hotWang,aplKakeya,TBenseman,TMinami}
In addition, the formation of such a hot spot has been explained by numerically solving the thermal diffusion equation for IJJs.~\cite{hotYurgens,hotGross}
Hence, temperature inhomogeneities such as the hot spots have been considered to play a crucial role in strong THz emission.
Since the critical current density $j_c$ depends on the temperature $T$,  the hot spot naturally induces an inhomogeneous $j_c$ distribution that strongly excites the Josephson plasma wave inside IJJs.\cite{inphaseKoshelev,inphaseAsai}
Therefore, local heating of IJJs by external heat sources, e.g., laser irradiation or current injection, is expected to be a promising method to realize intense THz emission.

In the present work,
a method for obtaining intense THz emission from mesa-structured IJJs using external local heating is theoretically proposed (see Fig.~1). 
The IJJ mesa is fabricated on an IJJ base crystal, and the mesa is covered by an upper electrode.
A focused laser beam is irradiated on the upper electrode and locally increases the temperature of the IJJ mesa beneath the electrode. 
We systematically investigate the THz emission of such a setup by solving the sine--Gordon, Maxwell, and thermal diffusion equations simultaneously. We report that the emission power is dramatically enhanced by local heating.
Based on the above analyses, we clarify the optimum heating conditions for achieving emission of over 1 mW of power.

\begin{figure}[b]
\includegraphics[width = 7.25cm]{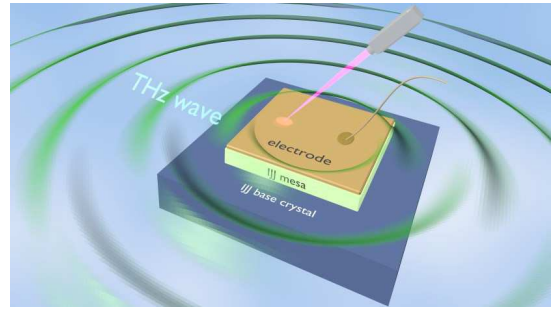}
\caption{Schematic of intrinsic Josephson junction (IJJ) mesa that is locally heated by laser irradiation.
The IJJ mesa is fabricated on an IJJ base crystal and covered by a metallic electrode.
A laser beam is incident on the upper electrode in order to locally increase the temperature of the mesa beneath the electrode. 
This system provides dramatic enhancement of THz emission power.}
\label{f1}
\end{figure}

\begin{figure}
\includegraphics[width = 7cm]{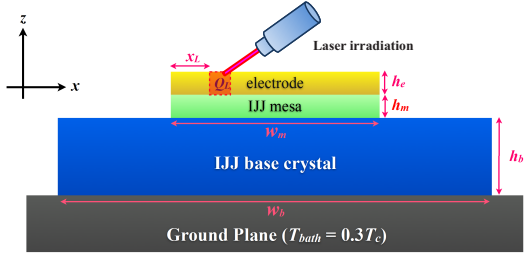}
\caption{ Two-dimensional model of our proposed IJJ emitter. The notation $x_L$ indicates the position of the heating spot. 
Here, $w_m = 60~\mu$m  is the width of the electrode and the IJJ mesa, $w_b = 220~\mu$m  is the width of the IJJ base crystal,   $h_e = 1~\mu$m is the thickness of the electrode,  $h_m = 1~\mu$m is the thickness of the IJJ mesa, and $h_b = 20~\mu$m  is the thickness of the IJJ base crystal. 
The DC voltage is applied to the IJJ mesa region. 
The external heating power is designated as $Q_L$.
}
\label{f2}
\end{figure}

In the IJJ THz emitter shown in Fig.~1, the local temperature of the IJJ mesa is controlled by laser heating.
An artificial $j_c$ distribution created by the local heating strongly enhances the THz emission.
To validate our concept and investigate the optimum conditions for intense emission, we perform a numerical simulation based on a two-dimensional model (see Fig.~2).   
The dimensions of the model are as follows: the width of the electrode and the IJJ mesa $w_m = 60~ \mu$m, the width of the IJJ base crystal $w_b = 220~\mu$m, the thickness of the electrode $h_e =1~\mu$m,
the thickness of the IJJ mesa $h_m =1~\mu$m and the thickness of the IJJ base crystal $h_b = 20~\mu$m.\cite{Ozyuzer,jpsjKado,hotWang,hot2Wang,aplKakeya,An,TBenseman,Sekimoto,TMinami}
A DC voltage is applied to the IJJ mesa region, and the base crystal is attached to an infinite ground plane.
In this study, we assume that the temperature distribution is not affected by the phase dynamics, and we solve the thermal diffusion equation upon assuming the Joule heating in the IJJ mesa.
Subsequently, we solve the sin--Gordon equation based on the obtained temperature distribution.

The thermal diffusion equation in IJJs is described as follows:
\begin{eqnarray}
0 = \frac{\partial}{\partial x} \Bigl[ \kappa_{ab}(T) \frac{\partial T}{\partial x} \Bigr] + \frac{\partial}{\partial z} \Bigl[ \kappa_c(T) \frac{\partial T}{\partial z}  \Bigr] 
  +  \frac{ j_{ex}^2}{\sigma_c(T)}.
\end{eqnarray}
Here, $T$ denotes the temperature, $\sigma_c$ denote the $c$-axis conductivity in the IJJ mesa, $j_{ex}$ denotes the external current density injected into the IJJ mesa region, and $\kappa_{ab} (\kappa_c)$ denotes the thermal conductivity along the $ab$ plane ($c$ axis). 
In the calculation, we include a heat source $Q_{L}$ in the electrode region that is indicated by the red shaded area in Fig.~2 in order to model the local heating by an external energy source such as laser irradiation.
 In this study, we assume that the spot size of $Q_{L}$ is 5~$\mu$m ($\ll~w_m$), and the position of the heating spot is defined by the notation $x_{L}$.   
 The enhancement of emission via local heating occurs as long as the spot size is smaller than the mesa size.
We use the temperature-dependent parameters $\kappa_{ab}$, $\kappa_c$ and $\sigma_c$ adopted in a previous theoretical study.\cite{hotGross}
In the thermal diffusion equation for upper electrode, we use the isotropic diffusion constant $\kappa = 20$ W/${\textrm m} \cdot {\textrm K}$.\cite{kappa} 
 We impose the boundary condition $T = T_{bath}$ at the boundary between the IJJ base crystal and the ground plane.
 In this study, we set the bath temperature to $T_{bath}= 0.3~T_c $.
 %
We confirm that the enhancement of THz emission discussed below is also observed for different values of $T_{bath}<T_c$.
  The open boundary condition $\nabla T = 0$ is used  for other boundaries.\cite{heatleak}

The dynamics of the phase differences $\phi$ in the IJJ mesa are described by the sine-Gordon equation within the in-phase approximation where all phase differences between the IJJ layers are equal to the common phase difference $\phi$ as,\cite{fdtdKoyama1,inphaseAsai}
\begin{eqnarray}
\frac{ \epsilon_c \hbar }{2 e d} \frac{\partial^2 \phi}{\partial t^2}   =   c^2 \frac{\partial B_y}{\partial x} - \frac{1}{\epsilon_0 }  \left[  j_c (T) \sin{\phi} + \sigma_c (T) E_z -j_{ex} \right] ,
\end{eqnarray}
where  $j_c(T)$ denotes the critical current density, $\epsilon_0$ and $\epsilon_c$ denote the vacuum permittivity and relative permittivity of the junctions, $d$ denotes the thickness of the insulating layers of the IJJs, $c$ denotes the light velocity, and $\Phi_0$ denotes the flux quantum.
The electromagnetic (EM) fields in the IJJs are given by $E_{z} (x,t) = \frac{\hbar}{2ed}  \frac{\partial \phi}{\partial t}$, $B_{y} (x,t) = \frac{\hbar}{2ed}\frac{\partial \phi}{\partial x}$.
We use the Ambegaokar--Baratoff relation  $j_c(T)/j_c(0) = \Delta(T)/\Delta(0) \tanh{(\Delta(T)/k_B T)}$ in order to calculate the $T$ dependence of  $j_c$, where $\Delta(T)$ denotes the BCS superconducting gap.\cite{reviewYTanaka}
In this study, we assume $\epsilon_c = 17.64$,\cite{jpsjKado}  $d = 1.2$ nm,\cite{FukuyaTachiki} and $j_c(0) = 4\cdot 10^2$ A/${\textrm {cm}}^2$.\cite{Inomata}

\begin{figure}[b]
\includegraphics[width = 7cm]{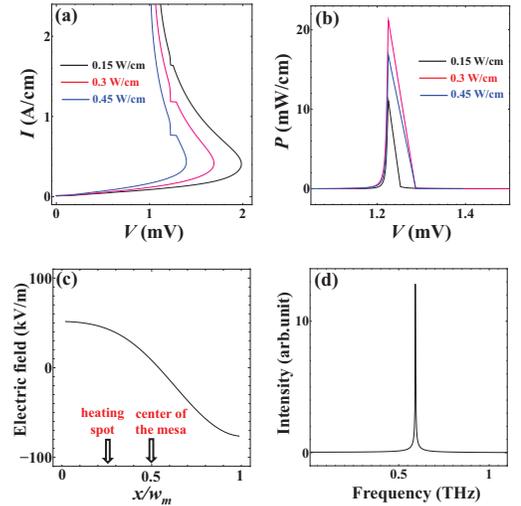}
\caption{Calculated results for the fixed heating-spot position of $x_{L} =13.3$ $\mu \mathrm{m}$. (a) Current $I$ versus voltage $V$ curves, and (b) emission power $P$ versus voltage $V$ curves for $Q_{L} = 0.15, 0.3,$ and $0.45~$W/cm.
(c) Snapshot of the oscillating part of electric field in the mesa, and (d) frequency spectrum of emission from the mesa at $I=1.19$~A/cm, $V=1.23$~mV for the case of $Q_{L} = 0.3$~W/cm. 
}
\label{f3}
\end{figure}


Based on the above numerical method, we calculate the THz emission power $P$ by varying the heating power $Q_{L}$ and the heating-spot position $x_{L}$.
First, for clarifying the optimum heating power, we vary $Q_{L}$ from 0 to 0.55 W/cm for a fixed heating-spot position of $x_{L} =13.3$~$\mu \mathrm{m} < w_m/2$.
In this calculation, we decrease the external current from the critical current value of the mesa 2.4~A/cm ($=j_{c} \times w_m$) to 0~A/cm and calculate $P$ for each current.
Figures 3 (a) and (b) show the current $I$ and the emission power $P$ as functions of the voltage $V$ for different values of heating power $Q_{L}$. 
In the plots, the voltage is divided by the number of the IJJ layer in order to indicate the voltage applied to each IJJ layer.
The $I$ vs.~$V$ curve exhibits a negative differential resistance.
This back-bending feature originates from strong self-heating in the high-current region (0.3--2.4 A/cm).\cite{Kurter}
The resistivity of the IJJs, $1/\sigma_c$, decreases with increase in the mesa temperature \cite{hotYurgens,hotGross}, and thus, the voltage across the IJJs is suppressed in the high-current region.
Importantly, in addition to the self-heating,  the local heating $Q_{L}$ increases the mesa temperature and decreases the voltage. Hence, the $I$ vs.~$V$ curves shift towards the low-voltage region with increase in $Q_{L}$ as shown in Fig.~3(a).
At $V=1.23$ mV, the $I$ vs.~$V$  curve exhibits a small kink, and the $P$ vs.~$V$ curve exhibits a sharp peak which indicates strong THz emission.
In order to clarify the origin of the intense emission, we calculate the distribution of the EM field inside the IJJ mesa.
 A snapshot of the oscillating part of the electric field at $I=1.19$~A/cm, $V=1.23$~mV for $Q_{L} = 0.3$~W/cm is shown in Fig.~3(c).
 As can be seen in the figure, a standing wave mode whose half-wavelength is equal to the mesa width $w_m$ appears.
Figure 3(d) shows the frequency spectrum of the EM wave emitted by the mesa. 
It is seen that a sharp peak appears around 0.59 THz, which is equal to the AC Josephson frequency $f_J =  2eV/h$ and the cavity resonance frequency given by $f_c = cn/(2 \hspace{-0.8mm}\sqrt{\epsilon_c}w_m)$, where $n = 1$.
In this manner, a strong THz wave is emitted from the mesa corresponding to the appearance of the cavity resonant EM mode.
We note that the intense emission at $V=1.23$ mV does not appear for $Q_{L} > 0.55$ W/cm because the strong external heating always ensures that the voltage across the IJJs less than the resonant voltage $V =1.23$ mV.


Figure~4(a) shows the plot of the emission power as a function of $Q_{L}$; this plot is used to examine the optimum heating power for practical THz emitters.
Remarkably, the emission power $P$ is dramatically enhanced by the external heat source $Q_L$ in comparison to the case without the external local heating, as observed from this figure. Our results indicate that the use of external local heating is a powerful method to achieve high-power THz emission. We have also confirmed that this enhancement is also observed for emission corresponding to other cavity resonance modes, e.g., for the mode $n=2$. 

It is particular noteworthy that the strongest emission is obtained around $Q_L =0.3$~W/cm. 
In order to study this behavior, we plot the spatial distribution of $T$ and $j_c$ in the mesa during the intense emission, as shown in Figs.~4(b) and (c), respectively.\cite{IVshift}
In the case of $Q_L =0.3$~W/cm, the heating-spot temperature is slightly lower than $T_c$, and thus a hot spot ($T>T_c$) is not formed.
The change in $j_c$ becomes significant when $T$ is slightly below $T_c$ as expected from the temperature dependence of $j_c$.\cite{reviewYTanaka}
Therefore, the drastic $j_c$ modulation via local heating strongly excites the THz Josephson plasma wave inside the IJJ mesa. 
On the other hand, for $Q_L < 0.3$~W/cm, the hot-spot region is formed during THz emission. 
Since the hot-spot region does not contribute to the emission because $j_c = 0$  in this region,
the formation of the hot spot results in reduction in the emission power.
Conversely, for  $Q_L > 0.3$~W/cm, the mesa temperature becomes considerably lower than $T_c$.  
%
In this case, the $j_c$ modulation becomes small compared to the case of $T \approx T_c$ because $j_c$ is nearly constant unless $T$ is close to $T_c$.
Consequently, the excitation of the THz Josephson plasma wave becomes weak.
Therefore, we can conclude that {\it the local heating that maintains the temperature of the heating spot slightly lower than $T_c$ is preferable for high-power emission}.
Importantly, our result is consistent with a recent experimental study that reported the strongest emission when $T$ was slightly lower than $T_c$.\cite{TBenseman}


\begin{figure}
\includegraphics[width = 7cm]{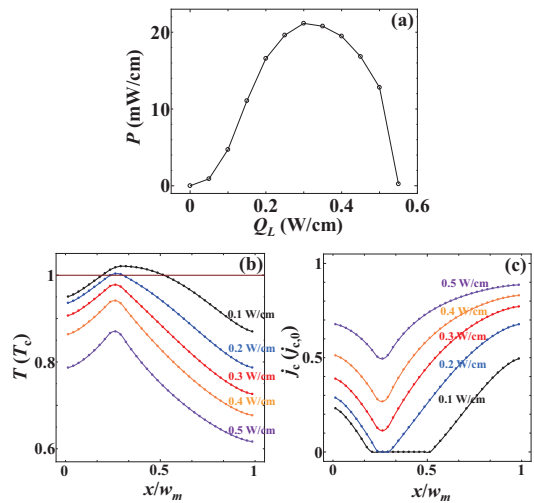}
\caption{Calculated results for fixed heating-spot position of $x_{L} =13.3 \mu \mathrm{m}$.  (a) Emission power $P$ as a function of $Q_{L}$.  (b) and (c) Distribution of $T$ and $j_c$ respectively, in the mesa for $Q_{L} = 0.1, 0.2, 0.3, 0.4$ and $0.5$ W/cm.}
\label{f1}
\end{figure}

\begin{figure}
\includegraphics[width = 7cm]{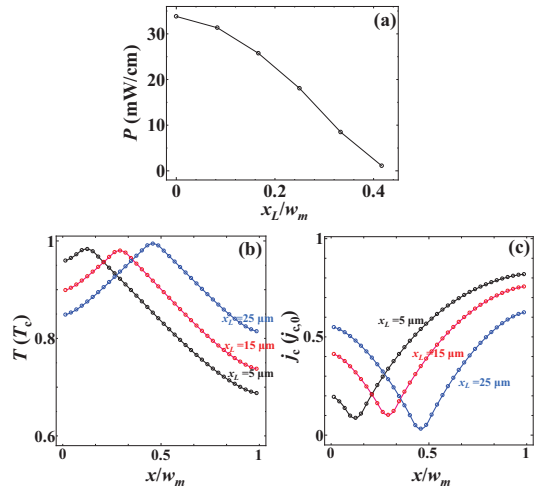}
\caption{Calculated results for the fixed heating power $Q_{L} =0.3$ (W/cm). (a) The emission power $P$ as a function of $x_{L}$. The distribution of (b) $T$ and (c) $j_c$  for different values of $x_{L}$.}
\label{f1}
\end{figure}

 Next, for clarifying the optimum heating-spot position,  we calculate the emission power by varying  $x_{L}$.
 We assume $Q_{L} = 0.3$ W/cm, which yields the highest emission power in the above calculation.  
Figure 5(a) shows the emission power as a function of $x_{L}$.
From the figure, we note that the emission power increases as the heating spot approaches the edge of the mesa.
The maximum emission power at $x_{L} =0$ $\mu$m  is  33.8 mW/cm.
The emission power of a 3D mesa is estimated by multiplying the emission power in the $x$-$z$ plane by the mesa length along the $y$ direction.
Consequently, the maximum emission power whose length is comparable to the experimental mesas $\sim~300 \mu$m \cite{Ozyuzer,jpsjKado,hotWang,hot2Wang,aplKakeya,An,TBenseman,Sekimoto,TMinami} reaches 1 mW.
 %
The distributions of $T$ and $j_c$ in the mesa during the intense emission are shown in Figs.~5(b) and (c), respectively.
Both distributions are asymmetric with respect to the center of the mesa ($x/w_m = 0.5$), and this asymmetric feature becomes more prominent with decrease in $x_{L}$.
Since the electric field corresponding to the $n=1$ cavity resonance mode is antisymmetric with respect to $x/w_m = 0.5$, as shown in Fig.~3(b),  
the $n=1$ mode is strongly excited by the asymmetric AC Josephson current.
Therefore, intense emission is obtained by heating the mesa edge because this leads to a large asymmetric $j_c$ distribution with respect to the mesa center.
It should be noted that the optimum position of the heating spot depends on the resonance mode in the mesa.
In the case of the $n=2$ cavity resonance mode, for instance, the electric field is symmetric with respect to $x/w_m = 0.5$.
Hence, for the $n=2$ mode, the local heating around  $x/w_m = 0.5$ is preferable for intense THz emission.

In conclusion, we have investigated THz emission from  IJJs  that are locally heated by an external heat source. 
We clarified the optimum heating condition to realize high power THz emission. 
The key points to design an intense THz emitter are 
(1) control of the heating power to maintain the mesa temperature slightly lower than $T_c$, and
(2) control of the heating position corresponding to the symmetry of the resonance modes in the mesa
(e.g. heating at the mesa edge is most preferable for the half-wavelength cavity-resonance mode.).
We have also investigated the above considerations for different bath temperatures, and we confirm that the above conditions for obtaining high-power emission do not depend on $T_{bath}$.
Recently, the temperature distribution in the mesa has been directly and precisely observed in experimental studies.\cite{TBenseman,TMinami}
Moreover, control of the spatial $T$ and $j_c$ distributions has been experimentally achieved in niobium-based Josephson junctions.\cite{Granata}  
In this light, we believe that the precise control of local temperature based on our theory will enable the practical realization of a high-power THz emitter using IJJs.

We wish to thank K. Kadowaki, I. Kakeya, T. Kashiwagi, H. Minami, F. Nori, Y. Ota, M. Tachiki, M. Tsujimoto, and H. B. Wang for fruitful discussions and comments. This work was supported by Grant-in-Aid for JSPS Fellows.


\begin{thebibliography}{00}
\bibitem{heatleak}  In this model, the heat leak through the upper electrode is not taken account. The importance of this leak has been pointed out in the previous experiments,\cite{aplKakeya} and it will reduce the mesa temperature. However, the heat flow through the electrode uniformly attached to the mesa only shift the averaged temperature of the mesa, and it will not affect the discussion in this paper.
\end{thebibliography}
\end{document}



\title{Supplemental Material: 

Intense terahertz emission from intrinsic Josephson junctions by external heat control}




\author{Hidehiro Asai}
\author{Shiro Kawabata}
\affiliation{Electronics and Photonics Research Institute (ESPRIT), National Institute of Advanced Industrial Science and Technology (AIST), Tsukuba, Ibaraki 305-8568, Japan}


\date{\today}

\pacs{}
\maketitle 


In this supplemental section, we discuss the effect of the variation in the thermal conductivity of the upper electrode $\kappa_e$ on the THz emission power.
%
In practical experiments, the electrode used is composed of good conductors such as gold, and the electrodes thickness $h_{Au}$ is $\sim$30 nm.\cite{aplKakeya,hotGross}
The thermal conductivity of thin-film gold, $\kappa_{Au}$, is 100--300~W/${\textrm m} \cdot {\textrm K}$.\cite{Aukappa}
On the other hand, in our calculation, we have assumed that $h_e$ is $1 \ \mu$m and $\kappa_e$ is $20$~W/${\textrm m} \cdot {\textrm K}$.
Although the values of $\kappa$ and $h$ in our calculation are different from those of actual experiments, both these parameters yield 
quantitatively the same amount of the heat transfer, as explained below.
%

The amount of the heat transfer along the width ($x$)~direction in the electrode is roughly given by the product of the cross section area ($\propto$ thickness, $h$) and the averaged heat flux ($\propto$ thermal conductivity, $\kappa$). 
Since $ h_{e} \times \kappa_e$  and $h_{Au} \times \kappa_{Au}$ are of the same order, the amounts of heat transfer through both electrodes are comparable.
On the other hand, the thermal conductivity of IJJs $\kappa_{ab} =$3--5 W/${\textrm m} \cdot {\textrm K}$~\cite{hotGross,IJJkappa} is considerably smaller than $\kappa_e$, and thus, the temperature distribution in the IJJ mesa largely depends on the heat transfer through the electrodes. Hence, the temperature distributions in our calculation are expected to be similar to those in experiments. 
%
%
%

%
\renewcommand{\figurename}{Fig S\hspace{-0.125cm}}
\begin{figure}[htbp]
\includegraphics[width = 7.5cm]{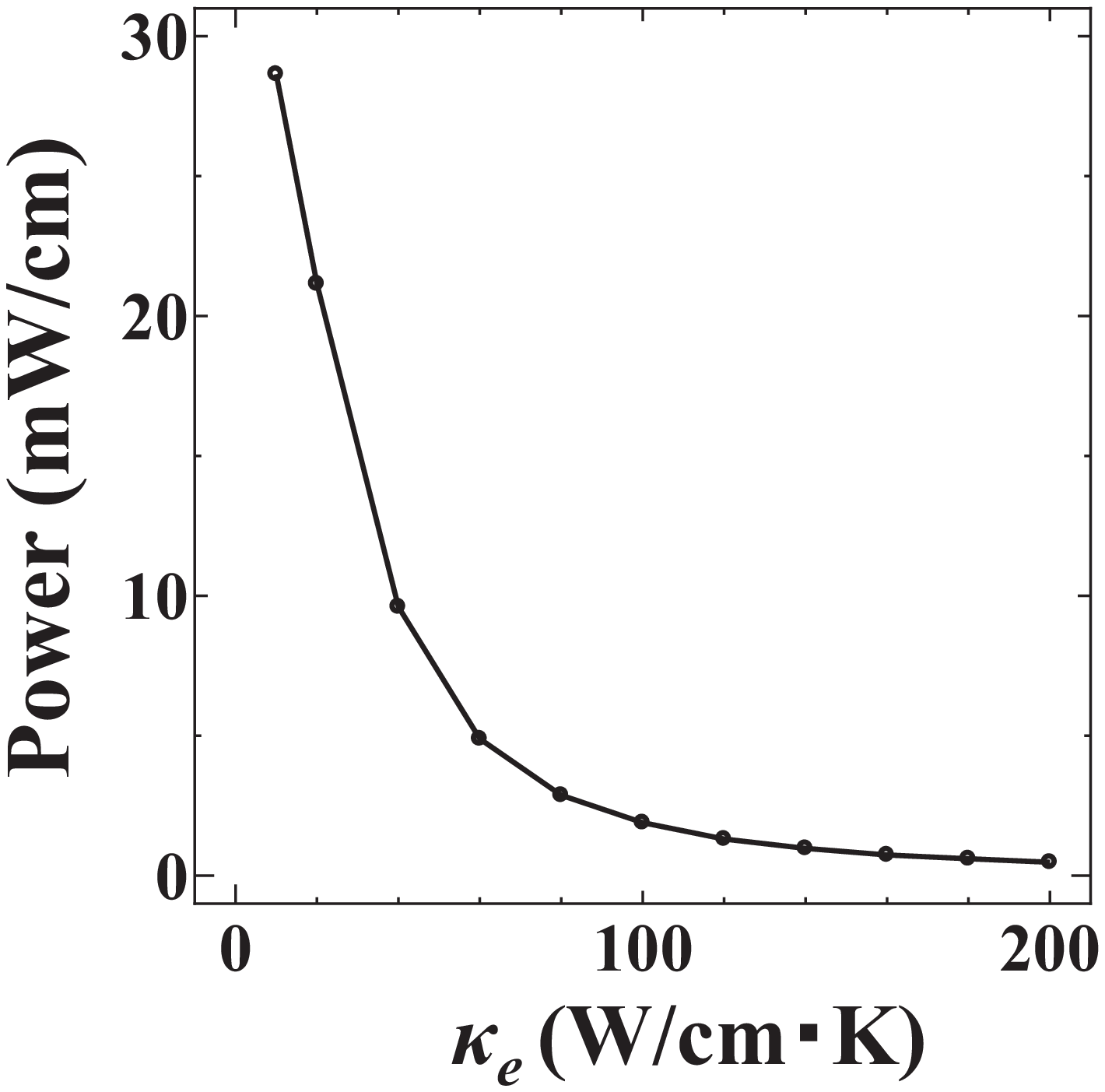}
\caption{Emission power as a function of thermal conductivity $\kappa_e$ for the case of $x_{L} =13.3$ $\mu \mathrm{m}$ and $Q_{L} = 0.3$ W/cm.}
\label{f1}
\end{figure}

Next, we calculate the emission power for varying values of $\kappa_e$. 
In this calculation, we set the heating-spot position $x_{L} =13.3$ $\mu \mathrm{m}$ and the heating power $Q_{L} = 0.3$ W/cm.
The dimensions of the system and other parameters are same as in our study.
Figure S1 shows the emission power as a function of $\kappa_e$.
From the figure, we note that the emission power decreases with increase in $\kappa_e$.
This behavior can be attributed to the change in the temperature distribution in IJJ mesas.
Figure S2 shows the temperature distribution in the mesa during THz emission for $\kappa_e =$ 40, 80 and 120~W/${\textrm m} \cdot {\textrm K}$. 
As $\kappa_e$ increases,  the localized heat induced by laser irradiation tends to spread uniformly over the electrode. 
Hence, the temperature distribution in the IJJ mesa that is located beneath the electrode becomes uniform with increase in $\kappa_e$ as shown in Fig. S2.
Therefore, the spatial modulation of the critical current density $j_c$  in the mesa becomes weak, and the excitation of Josephson plasma wave is suppressed.

\begin{figure}[htbp]
\includegraphics[width = 7.5cm]{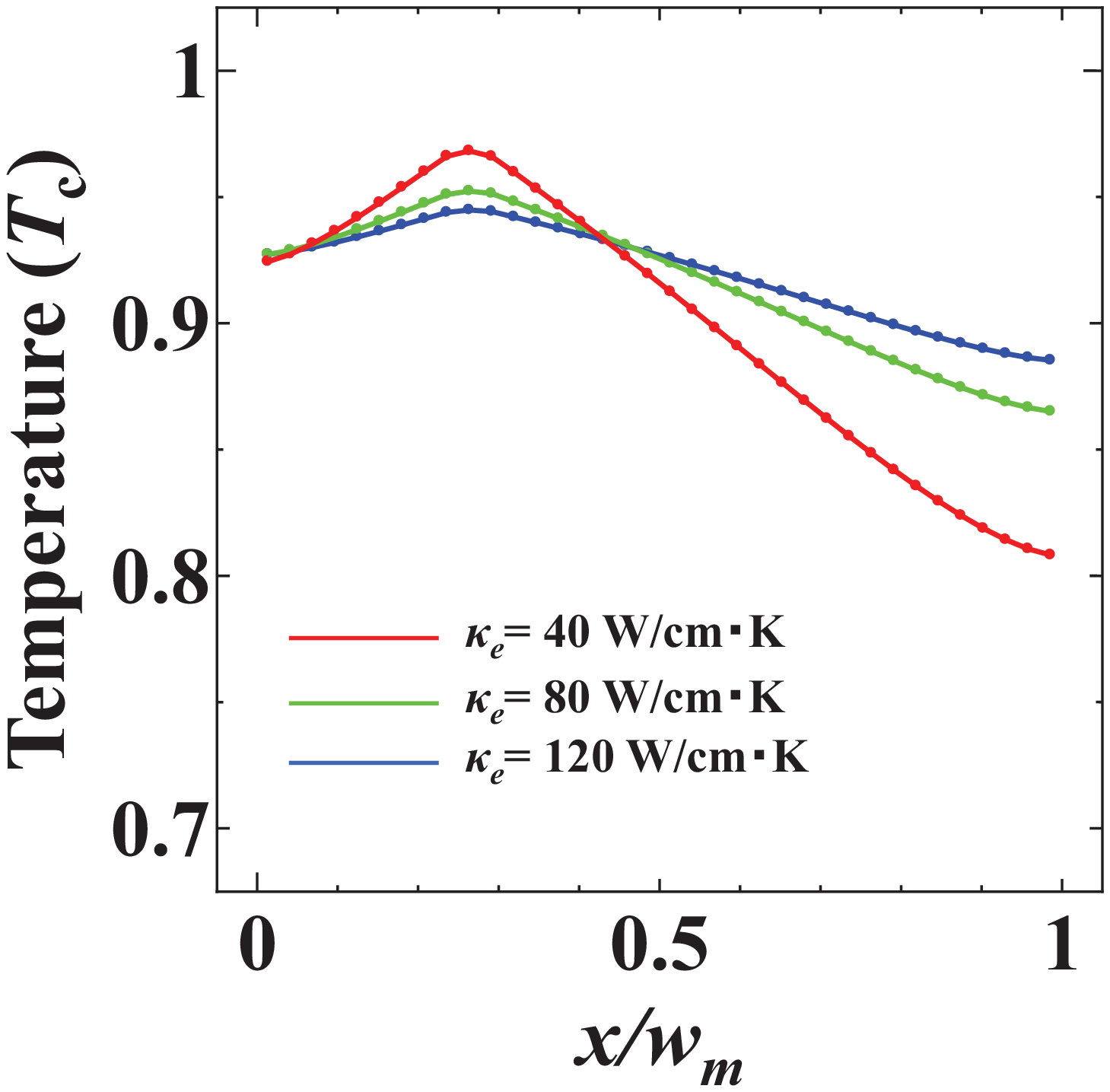}
\caption{Temperature distribution in the mesa for $\kappa_e =$ 40, 80 and 120~W/${\textrm m} \cdot {\textrm K}$. }
\label{f1}
\end{figure}

In this manner,  the occurrence of large amount of heat transfer due to large $\kappa_e$ causes a uniform temperature distribution in the mesa and suppresses the emission power.
As discussed above, the amount of heat transfer through electrodes is proportional to the product of the electrode thickness and thermal conductivity.
Therefore, the dependence of THz emission power on the thickness $h_e$ is expected to be similar to that for $\kappa_e$; the THz emission power is expected to be suppressed with increase in $h_e$.
This peculiar behavior is consistent with recent experimental observations.~\cite{aplKakeya}
The authors of this study have investigated the THz emission from several IJJ mesas with different electrode thicknesses, and they report that THz emission disappears when electrodes with large thicknesses are used.\cite{aplKakeya} 

In summary, the THz emission power from an IJJ mesa strongly depends on the heat transfer thorough the upper electrodes.
The THz emission power increases as the thermal conductivity and the electrode thickness decrease.